\documentclass[a4paper,11pt]{article}

\usepackage{caption}
\usepackage{subcaption}
\usepackage{pos}
\usepackage{lipsum} 
\usepackage{wrapfig,lipsum,booktabs}
\usepackage{float}
\usepackage{todonotes}
\usepackage{array,multirow,graphicx}
\usepackage{wrapfig}
\usepackage{lineno}

\title{Sensitivity of the IceCube Upgrade to Atmospheric Neutrino Oscillations}

\ShortTitle{Sensitivity of the IceCube Upgrade to Atm. Neutrino Oscillations}

\author{The IceCube Collaboration \\{\normalsize \normalfont(a complete list of authors can be found at the end of the proceedings)}\\}

\emailAdd{philipp.eller@tum.de}
\emailAdd{kayla.leonard@icecube.wisc.edu}
\emailAdd{jcw5951@psu.edu}
\emailAdd{rasmus.orsoe@tum.de}

\abstract{
IceCube DeepCore, the existing low-energy extension of the IceCube Neutrino Observatory, was designed to lower the neutrino detection energy threshold to the GeV range. A new extension, called the IceCube Upgrade, will consist of seven additional strings installed within the DeepCore fiducial volume. The new modules will have spacings of about 20 m horizontally and 3 m vertically, compared to about 40-70 m horizontally and 7 m vertically in DeepCore. It will be deployed in the polar season of 2025/26. This additional hardware features new types of optical modules with multi-PMT configurations, as well as calibration devices. This upgrade will more than triple the number of PMT channels with respect to current IceCube, and will significantly enhance its capabilities in the GeV energy range. However, the increased channel count also poses new computational challenges for the event simulation, selection, and reconstruction. In this contribution we present updated oscillation sensitivities based on the latest advancements in simulation, event selection, and reconstruction techniques.

\vspace{4mm}
{\bfseries Corresponding authors:}
Philipp Eller$^{1*}$, Kayla Leonard DeHolton$^{2}$, Jan Weldert$^{2}$, Rasmus \O rs\o e$^{1}$\\
{$^{1}$ \itshape Technical University of Munich, TUM School of Natural Sciences, Physics Department, 85747 Garching, Germany}\\
{$^{2}$ \itshape Dept. of Physics, Pennsylvania State University, University Park, PA 16802, USA}\\[4mm]
$^*$ Presenter

\ConferenceLogo{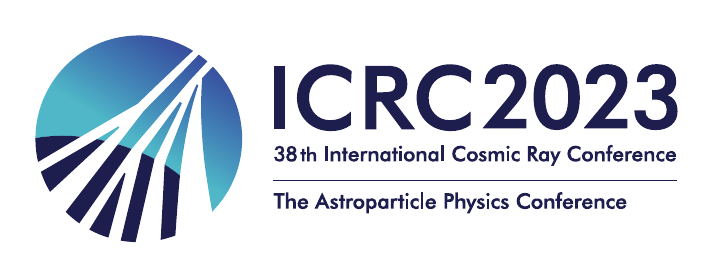}

\FullConference{The 38th International Cosmic Ray Conference (ICRC2023)\\ 26 July -- 3 August, 2023\\ Nagoya, Japan}
}

\begin{document}

\maketitle

\section{Introduction}
The IceCube Neutrino Observatory has been detecting atmospheric neutrinos in the GeV energy range using its low-energy extension DeepCore for the past decade \cite{Aartsen:2016icecube, IceCube:2011deepcore}. This data has been used to make precision measurements of the atmospheric neutrino oscillation parameters \cite{IceCube:2017numuprl, IceCube:2019nutauprd, IceCube:2023oscnextprd}
, tau neutrino appearance \cite{IceCube:2019nutauprd} and preference for the neutrino mass ordering (NMO) \cite{IceCube:2019dyb}.
A new extension, called the IceCube Upgrade, will be deployed in the polar season of 2025/26 and will consist of seven additional strings installed within the DeepCore fiducial volume. The new modules will have spacings of about 20\,m horizontally and 3\,m vertically, compared to about 40-70\,m horizontally and 7\,m vertically in DeepCore. This additional hardware features new types of optical modules with multi-PMT configurations. DEggs consist of two 8-inch PMTs pointing up and downards \cite{IceCube:2022degg}, while mDOMs consist of 24 3-inch PMTs pointing in all directions \cite{IceCube:2019anq}, as well as new calibration devices. This Upgrade will more than triple the number of PMT channels with respect to current IceCube, and will significantly enhance its capabilities in the GeV energy range.

The work presented here uses events simulated for the full 93-string detector configuration (``IC93'') using the state-of-the-art IceCube simulation chain, and new techniques that have been developed to clean the raw data, reconstruct events, and reject backgrounds (Sec.~\ref{sec:sim_proc}). We quantify the expected performance of the Upgrade detector extension, and compare to a scenario without the extension (Sec.~\ref{sec:results}). The sensitivities presented here supersede previous ones \cite{Ishihara:2019aao}.



\section{Simulation \& Processing}
\label{sec:sim_proc}

\subsection{Simulation}
\label{sec:sim}
Events are simulated following a similar procedure to Ref.~\cite{IceCube:2023oscnextprd}, including neutrino interactions with GENIE \cite{genie:2009} and the two primary backgrounds (atmospheric muons and pure detector noise triggers), assuming our latest ice model which takes into account birefreingence due to the crystalline structure of the ice \cite{Rongen:2021rgc}. For the IceCube and DeepCore strings, extra ice scattering and absorption in the refrozen boreholes is included due to different optical properties compared to the original glacial ice. For the Upgrade strings, the local ice property changes are not necessary because of a planned de-gassing procedure that will be used to eliminate air bubbles introduced in deployment. Event trigger and filtering is based on the current DeepCore implementation but adjusted to take into account the new multi-PMT Upgrade modules, delivering a realistic baseline performance, while further studies for optimized triggers and detector calibration with the new devices are ongoing.

\subsection{Noise Cleaning}
\label{sec:clean}
\begin{figure}[h]
    \centering
    \vspace{-12pt}
    \includegraphics[width=0.86\textwidth]{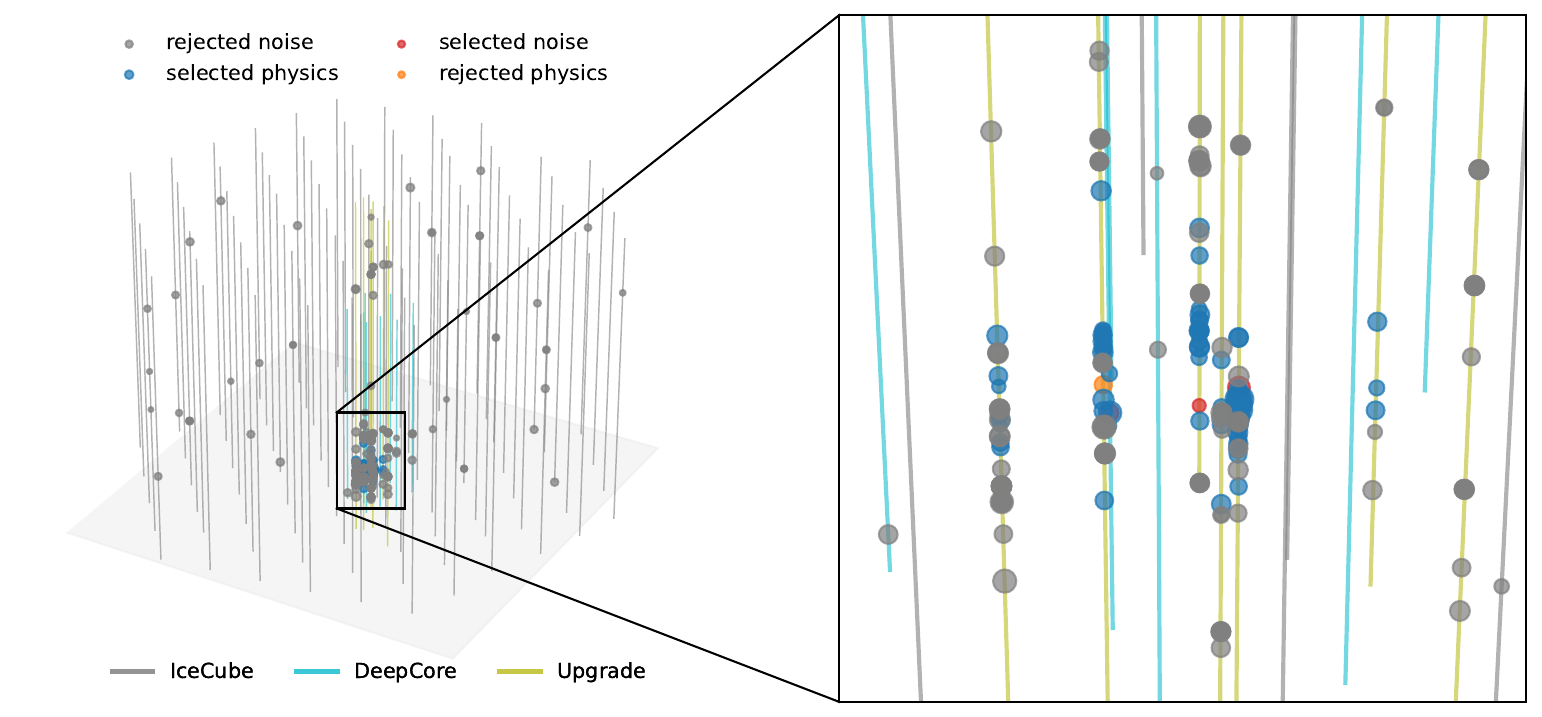}
    \caption{Event display of a 4.7 GeV muon neutrino interaction in the IceCube Upgrade. The existing IceCube and DeepCore strings are shown in grey and cyan, respectively, while the new IceCube Upgrade strings are shown in olive. This event has 96 hits caused by the neutrino interaction, of which 93 are correctly identified by the GNN (blue markers) and 3 are rejected (orange). The remaining 242 hits are caused by random detector noise: 240 are correctly rejected (grey) while 2 incorrectly pass the cleaning (red).}
    \label{fig:noise}
\end{figure}

Radioactive decays in the glass housing of the optical modules are the largest source of noise pollution in the data. In the past, it has been sufficient to use cuts on the temporal and spatial distance of hits for noise cleaning. However, because of the increased channel count and the increased levels of radioactivity in the new modules' glass, these cuts are no longer sufficient for the IceCube Upgrade.
We developed a new cleaning algorithm based on a Graph Neutral Network (GNN) specifically for the IceCube Upgrade. It performs a binary classification to predict whether a pulse is signal or noise. The GNN encodes the irregular grid of the detector and the irregular topology by representing each event as a point-cloud graph, where each node represents an observed pulse in the detector and edges are drawn to the 8 nearest hits. For each node, we use features that describe the observed pulse such as charge and arrival time, and features that describe the PMT such as its position, relative optical efficiency, direction, and surface area.

We train a customized version of the GNN DynEdge \cite{dynedge_paper} from the open-source framework \textit{GraphNeT} \cite{graphnet_paper}, using a dataset with about 4 million neutrino and muon events. We apply a threshold of 0.7 on the classification output. An example event cleaned by the GNN is shown in Fig. \ref{fig:noise}. The GNN reduces the average amount of noise per neutrino event by a factor of about 10 with only a minor loss in signal. Before cleaning, 70\% of hits in the uncleaned pulse series are noise, and after cleaning an average of 6.8\% of noise hits remain. In addition, when the GNN is applied to pure noise events, less than 1\% of these events survive. Neutrino events with at least 8 hits after cleaning on average retain 91\% of the signal.

\subsection{Reconstruction \& Classification}
\label{sec:reco}
\begin{figure}[h]
     \centering
     \vspace{-1em}
     \begin{subfigure}[b]{0.4\textwidth}
         \centering
         \includegraphics[width=1.\textwidth]{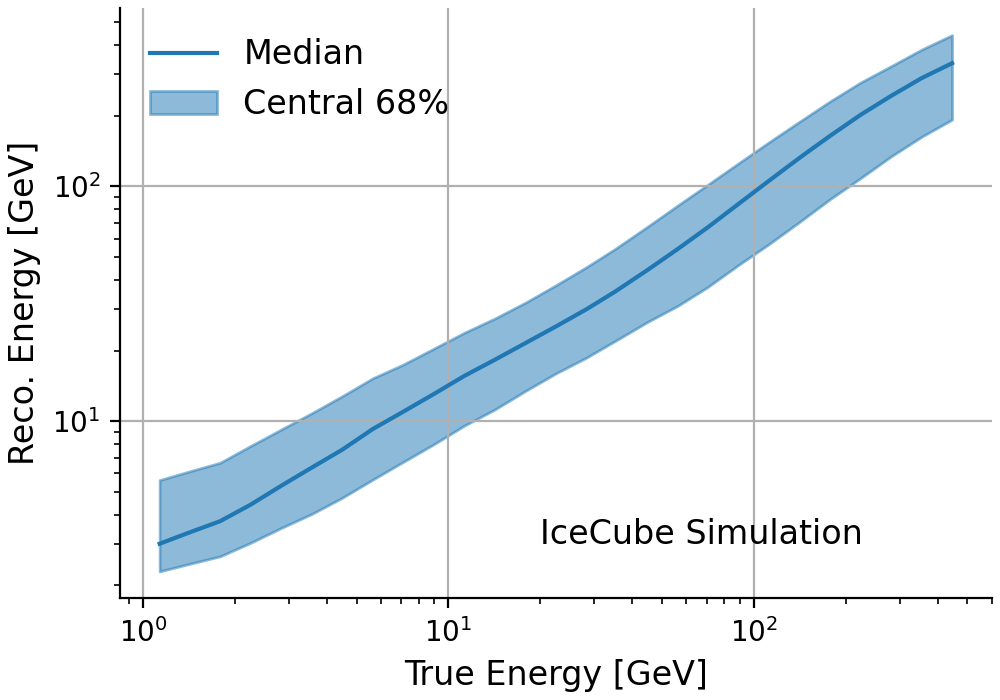}
         \caption{Energy}
         \label{fig:energy_reco}
     \end{subfigure}
     \hspace{0.03\textwidth}
     \begin{subfigure}[b]{0.4\textwidth}
         \centering
         \includegraphics[width=1.\textwidth]{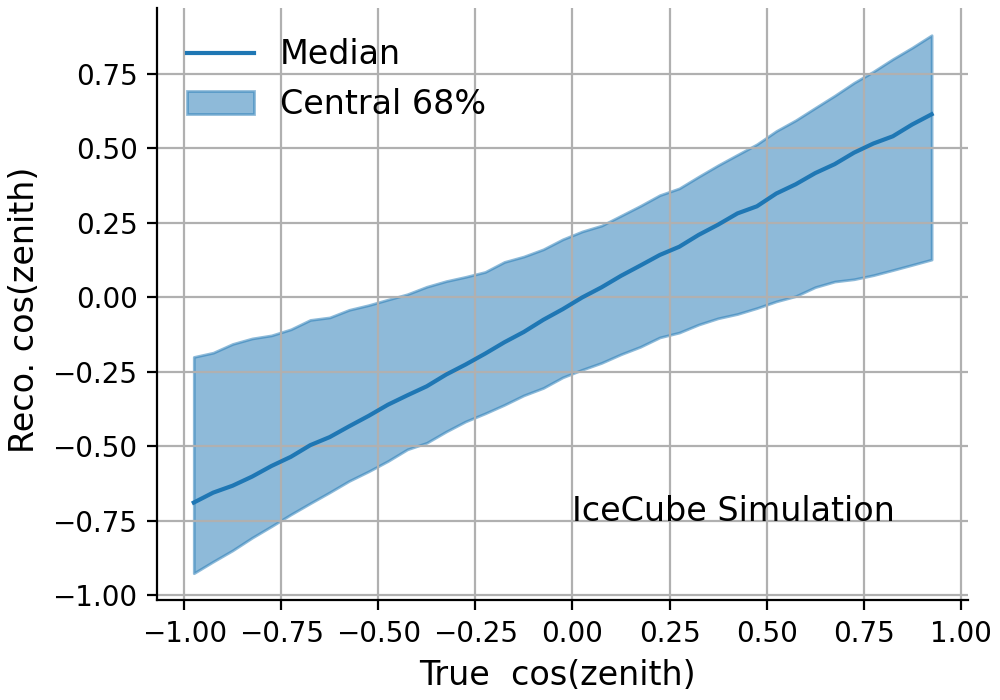}
         \caption{Zenith Angle}
         \label{fig:coszen_reco}
     \end{subfigure}
        \caption{Performance for the reconstruction of neutrino energy (left) and zenith angle (right). Shown are the median and central 68\% region of the reconstructed quantity as a function of the simulated truth. The bending of the angle towards the edges is due the quantity being bound $\in [-1,1]$.}
        \label{fig:reco}
\end{figure}

After events are cleaned, four separate instances of DynEdge are trained to reconstruct and classify events based on the cleaned pulse series. These tasks include reconstructing the neutrino energy, zenith angle \& uncertainty,  and classifying events by particle type (muon vs. neutrino) and event topology  (track vs. cascade). Each of the 4 models is trained on a different task-specific subsamples. Sample balancing is performed for the classification tasks. Figure \ref{fig:reco} shows the reconstruction performance for energy and cos(zenith). We find that the neutrino/muon classifier provides several orders of magnitude of background reduction. The track/cascade classifier achieves a 0.82 Area Under the Curve (AUC) score, which is a significant improvement compared to what's previously been reported for DynEdge for IC86 \cite{dynedge_paper}.

\subsection{Event Selection}
\label{sec:selection}
\begin{wrapfigure}{r}{0.47\textwidth}
    \centering
    \vspace{-1em}
    \includegraphics[width=0.47\textwidth]{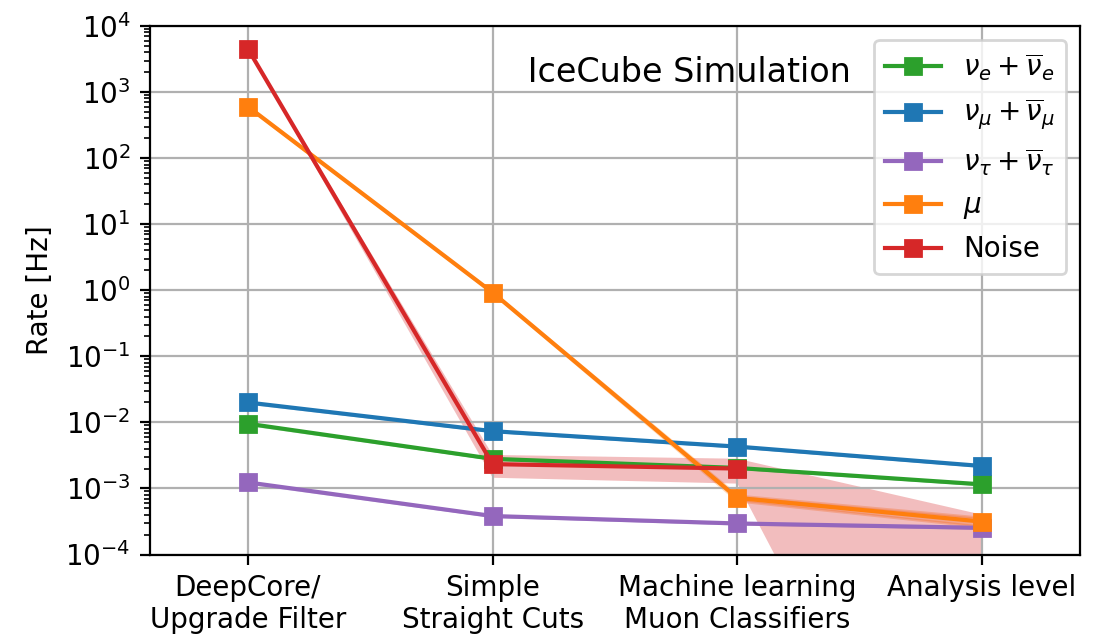}
    \caption{Event rates as a function of selection step, from filter level up to the analysis level.}
    \label{fig:rates}
\end{wrapfigure}

\begin{figure}[h]
     \centering
     \begin{subfigure}[b]{0.32\textwidth}
         \centering
         \includegraphics[width=\textwidth]{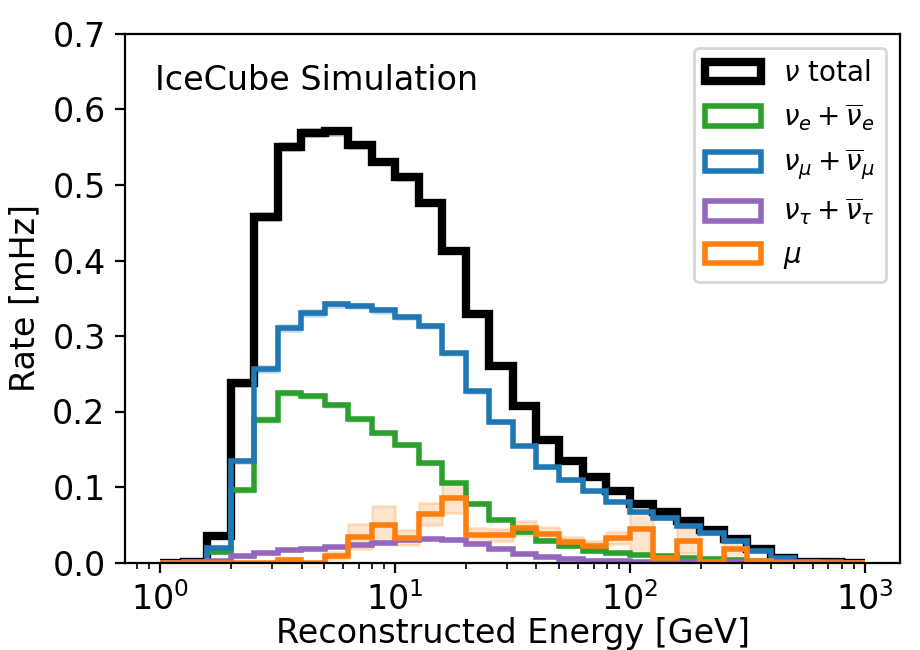}
         \caption{Energy}
         \label{fig:energy}
     \end{subfigure}
     \hfill
     \begin{subfigure}[b]{0.32\textwidth}
         \centering
         \includegraphics[width=\textwidth]{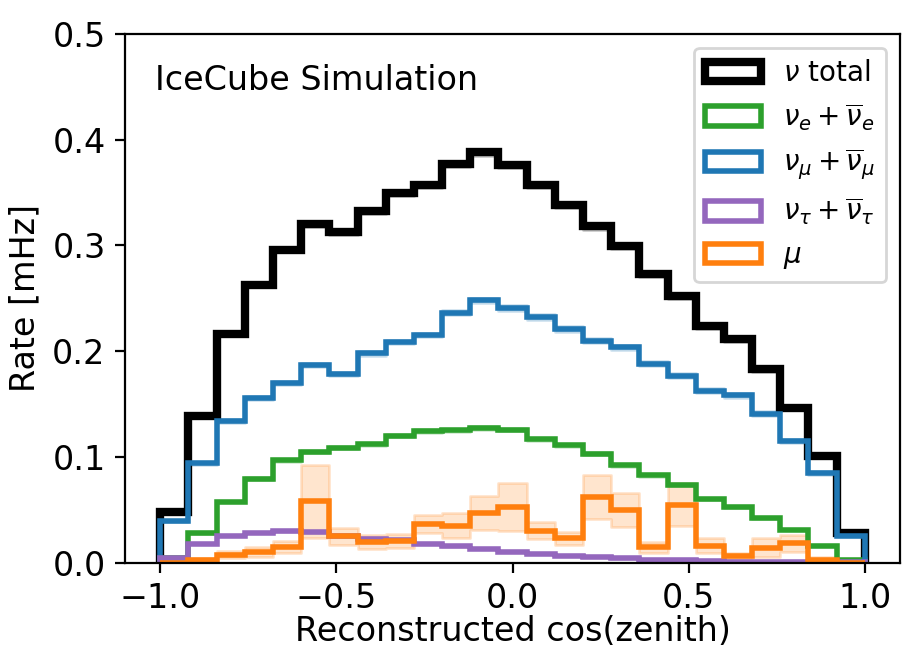}
         \caption{Zenith Angle}
         \label{fig:coszen}
     \end{subfigure}
     \hfill
     \begin{subfigure}[b]{0.32\textwidth}
         \centering
         \includegraphics[width=\textwidth]{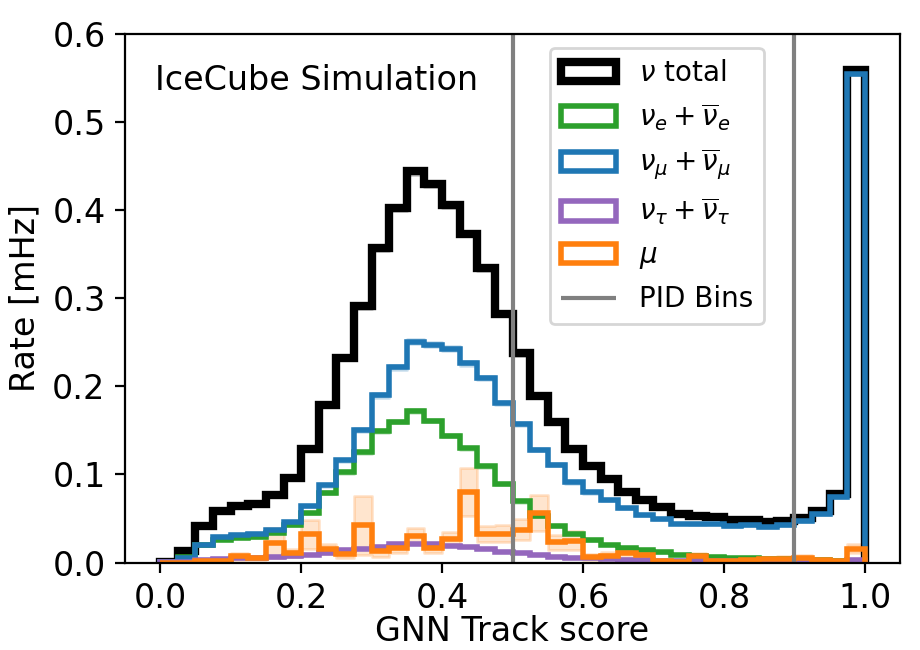}
         \caption{Event Type Classifier}
         \label{fig:pid}
     \end{subfigure}
        \caption{Distributions of final level event rates as a function of the three analysis observables. The shaded bands show $1\sigma$ statistical uncertainty, most visible for the muon background due to lower MC statistics.}
        \label{fig:final}
\end{figure}

Several levels of cuts are used to suppress the main backgrounds (atmospheric muons and noise) and achieve a high-purity neutrino sample, summarized in Fig.~\ref{fig:rates}. The first round consists of a few simple straight cuts, similar to what was used in Ref.~\cite{IceCube:2023oscnextprd}, that eliminate the most obvious background events.
The cut with the largest rejection power is the number of pulses 
remaining after the noise cleaning procedure described above. After the straight cuts, a collection of machine learning classifiers are used to identify muon background events. One of these is the muon/neutrino GNN classifier described above. There are also two Boosted Decision Trees (BDTs) that use reconstructed quantities and variables describing the spatial and temporal distribution of the hits. These three classifiers are able to suppress the remaining muon background by three orders of magnitude resulting in a sample that is heavily dominated by neutrinos. A few final cuts are made to focus on the the events for which most of the oscillation signal comes from. Figure \ref{fig:final} shows the final level event distributions after these cuts are made.

\section{Expected Oscillation Analysis Sensitivities}
\label{sec:results}

The analyses presented here largely follow previous DeepCore analyses, such as \cite{IceCube:2023oscnextprd}, and reflect realistic baseline sensitivities, while there is still room for improvement from better knowledge of systematic uncertainties and further optimization of the event selection and analysis choices. New, necessary parameters are added to take into account the new module types such as a new optical efficiency scale, and additional low energy cross section uncertainties are added since the energy threshold is lowered to 3\,GeV. This leads to a different treatment for (simulated) data coming before and after the new strings are deployed. Therefore, the different time periods are treated as separate data sources and we run a combined, simultaneous fit with correlated parameters.

We present the IceCube sensitivities to the atmospheric oscillation parameters ($\Delta m^2_{31}$ and $\theta_{23}$), a non-unitarity test with tau neutrinos, and the neutrino mass ordering.
We compare two scenarios: one with no new hardware installed and data-taking continuing as it is today (denoted as ``IC86'', reflecting the current 86 string detector). The other scenario includes the combined fit between 12 years of IC86 with additional years using the extra seven strings of the IceCube Upgrade from 2026 onwards (denoted as ``IC86 (12 yr) + IC93'').
Since the NMO signal is at a lower energy than the other analyses, different systematic uncertainties become more/less relevant, and therefore a slightly different set of parameters is used for the NMO analysis. An overview of the systematic uncertainties used is provided in Table~\ref{tab:sys}, and more details on each can be found in Ref.~\cite{IceCube:2023oscnextprd}. The largest contributions to uncertainty are currently from flux parameters, though this can be improved in the future by switching to newer flux models with reduced uncertainty \cite{daemonflux:2023}. The analyses assume true normal ordering and oscillation parameter values from NuFit 5.2 \cite{Esteban:2020cvm}.


\begin{table}[H]
\footnotesize
\begin{tabular}{l|l|ccc}
\textbf{Description} & \textbf{Parameter(s)}                                                                                                                                        & \textbf{Atm. osc.} & $\mathbf{\nu_\tau}$ & \textbf{NMO} \\ \hline
         
\textbf{Flux} & & & & \\
Spectral index  &     $\gamma$              &         x          &     x         &      x       \\
           \multirow{3}{7.5cm}{Uncertainty on  Pion and Kaon production (Barr et al. \cite{Barr_2006})} & $d_\pi$           & -                  & -              & x            \\
           & $g_\pi$, $h_\pi$, $i_\pi$, $w_K$, $z_K$ & x                  & x              & x            \\
           & $y_K$             & x                  & x              & -            \\
           Neutrino and Muon Normalizations                                                                                                                             & $ A_{eff} $, $\mu_{atm}$  & x                  & x              & x            \\ \hline
\textbf{Cross sections} & & & & \\
Deep inelastic scattering uncertainty \cite{NuGen}        & $\mathrm{DIS}_{\mathrm{CSMS}}$          & x                  & x              & -            \\
               Axial masses for Resonant CC and Quasi-elastic scattering                                                                                                 & $M_{A,res}^{CC}$, $M_{A,QE}$  & x                  & x              & x            \\
               Axial masses for Resonant NC and Coherent $\pi$ scattering                                                                                                  & $M_{A,res}^{NC}$, $M_{A,coh}$  & -                  & -              & x            \\
               Model uncertainty on tau neutrino cross section \cite{Conrad:2010mh}                                                                                 &      $\nu_{\tau}$ xsec            & -                  & -              & x            \\ \hline
\textbf{Detector} & & & & \\
Bulk ice properties scattering and absorption                                                                                                     &    scat., abs.               & x                  & x              & x            \\
               Optical module efficiencies (IceCube and Upgrade modules)                                                                                          & $\mathrm{OM}_\mathrm{eff, ICDC}$, $\mathrm{OM}_\mathrm{eff, ICU}$   & x                  & x              & x            \\
               Angular acceptance (IC86 configuration only)                                                                                                              &        $p_0, p_1$           & x                  & x              & x            \\ \hline
\textbf{Oscillations} & & & & \\
\multirow{2}{*}{Mixing Angles}      &  $\theta_{13}$                 & -                  & -              & x            \\
               & $\theta_{23}$                         & M                  & x              & x            \\
              Mass splitting      &     $\Delta m^2_{31}$              & M                  & x              & x            \\
              Unitarity breaking parameter         &    $\nu_\tau$ norm               & -                  & M              & -            \\
              Neutrino Mass Ordering                                                                                                                                      &      NMO             & NO                 & NO             & M           
\end{tabular}
\caption{Configuration of nuisance parameters for the analyses presented. The marker ``x'' indicates that a systematic is included in the fit, ``-'' means not included, and ``M'' denotes the measurement variable(s).}
\label{tab:sys}
\end{table}

\subsection{Atmospheric Oscillation Parameters}

\begin{figure}[h]
     \centering
     \begin{subfigure}[t]{0.36\textwidth}
         \centering
         \includegraphics[width=1.\textwidth]{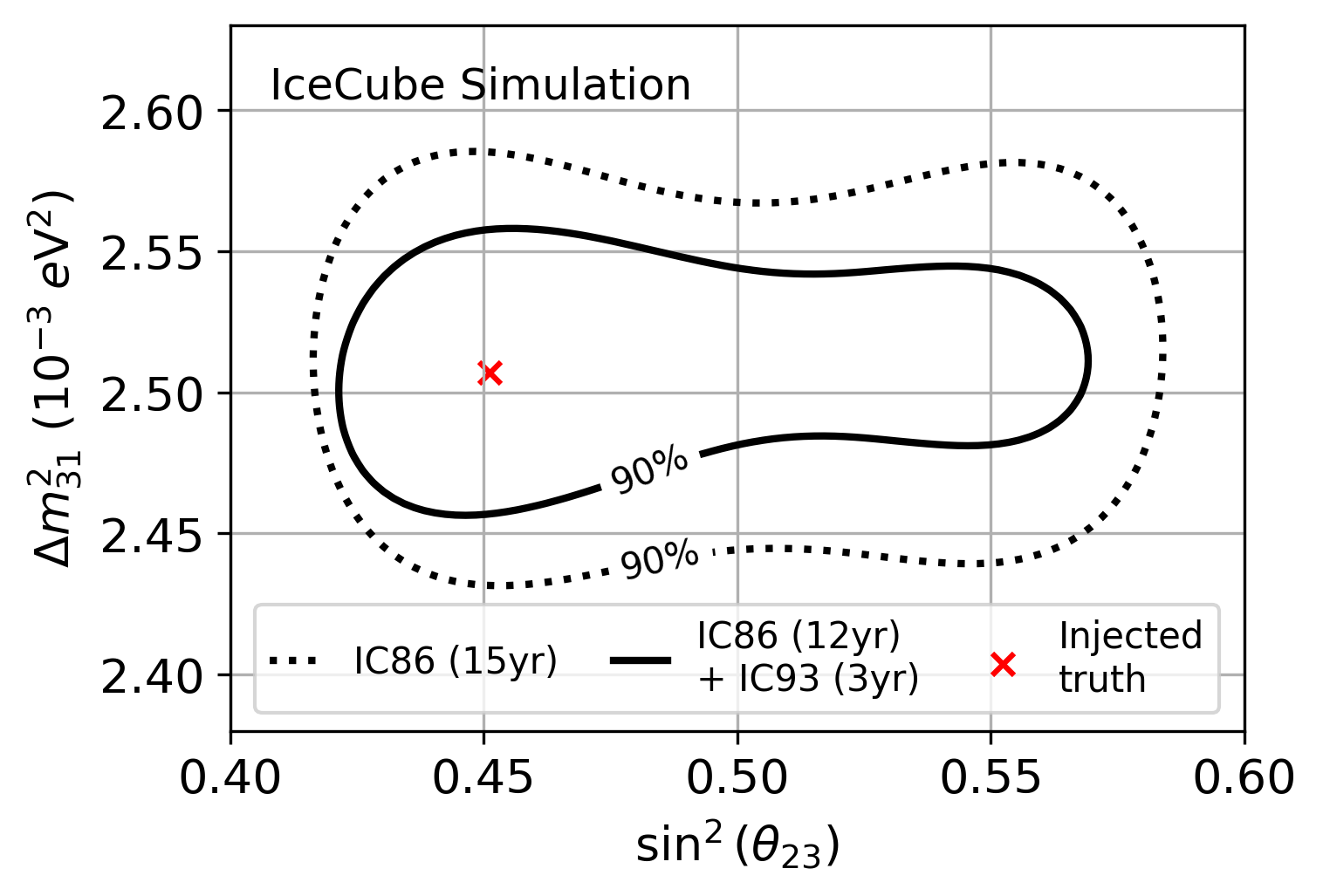}
         \caption{Joint 90\% CL sensitivity contours for the mixing angle $\theta_{23}$ and mass splitting $\Delta m^2_{31}$.}
         \label{fig:numu_app_2d}
     \end{subfigure}
     \hfill
     \begin{subfigure}[t]{0.62\textwidth}
         \centering
         \includegraphics[width=1.\textwidth]{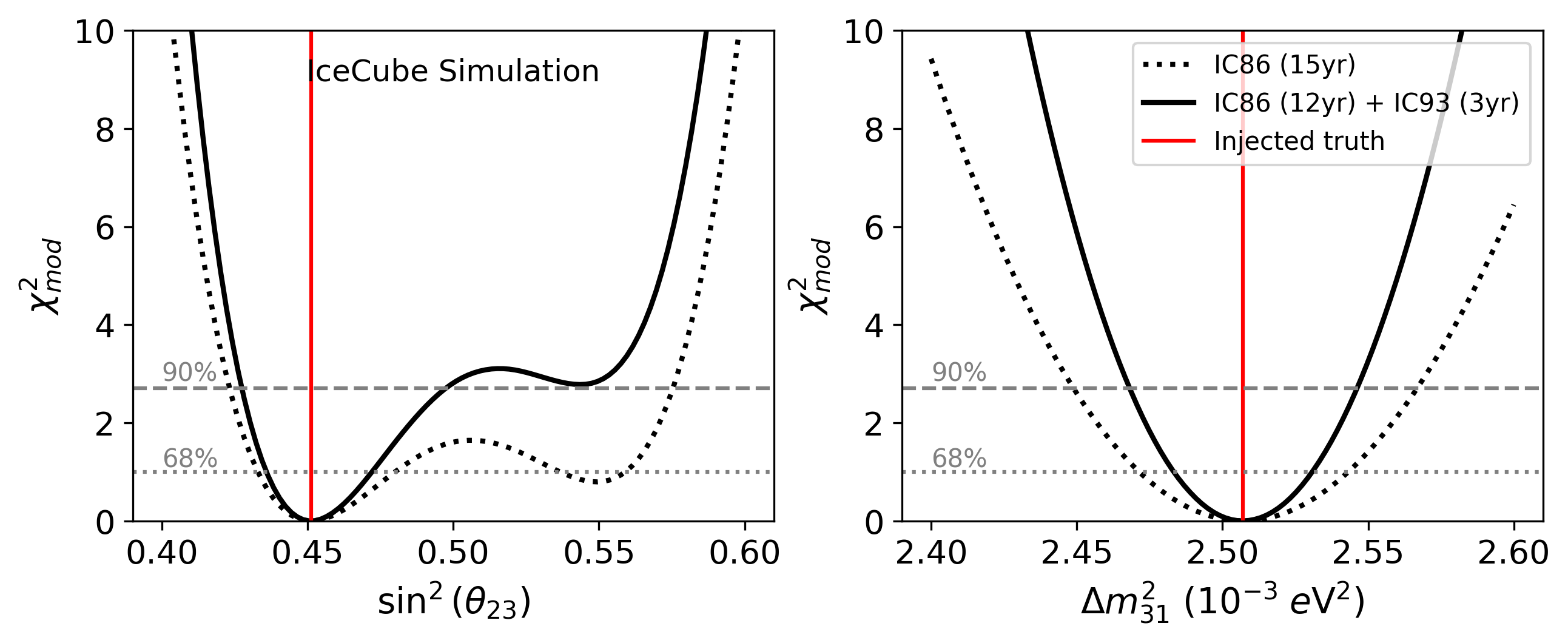}
         \caption{One dimensional profile $\Delta \chi^2$ curves separately for both measurement parameters}
         \label{fig:numu_app_1d}
     \end{subfigure}
        \caption{Sensitivity contours and profiles, respectively, for the standard atmospheric mixing analysis. The solid black lines show the scenario where the IceCube Upgrade is in place, while the dashed lines show the performance without additional hardware. The assumed true value is indicated in red.}
        \label{fig:numu_app}
\end{figure}

We follow a similar analysis procedure to existing IceCube measurements of the atmospheric oscillation parameters \cite{IceCube:2023oscnextprd}. Figure~\ref{fig:numu_app_2d} shows the sensitivity at the 90\% confidence level after 3 years with the new strings. In Fig.~\ref{fig:numu_app_1d}, one dimensional projections to the oscillation parameters are shown. The new strings increase IceCube's sensitivity to $\Delta m^2_{31}$ and $\theta_{23}$ by about 20-30\% and allow for a significantly better constraint of the atmospheric neutrino oscillation parameters.

\subsection{Non-unitary Mixing: Tau Neutrinos}

\begin{figure}[h]
     \centering
     \begin{subfigure}[t]{0.35\textwidth}
         \centering
         \includegraphics[width=1.\textwidth]{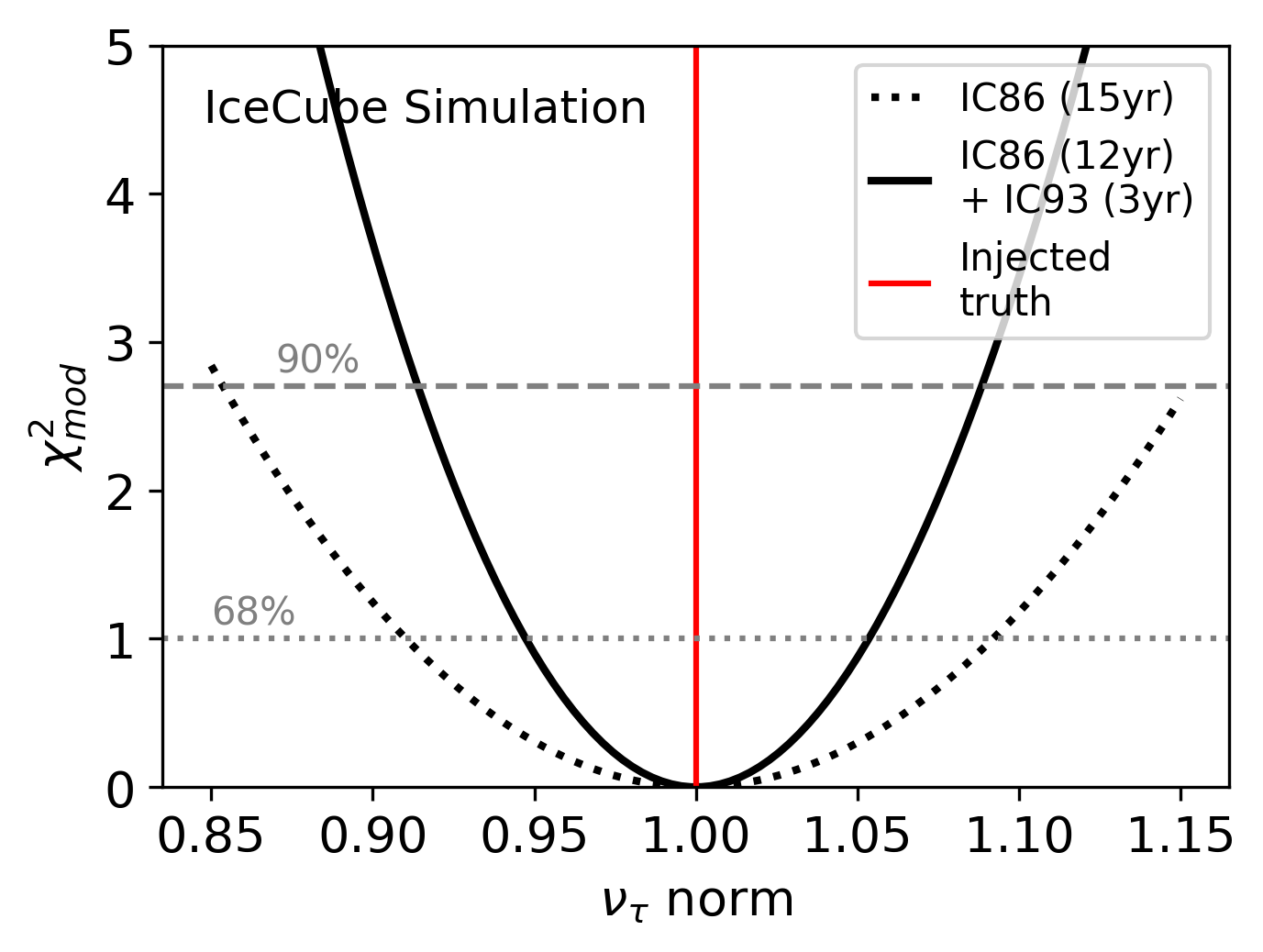}
         \caption{Profile $\Delta \chi^2$ curves for 15 years of livetime with and without including Upgrade}
         \label{fig:nutau}
     \end{subfigure}
     \hspace{0.03\textwidth}
     \begin{subfigure}[t]{0.4\textwidth}
         \centering
         \includegraphics[width=1.\textwidth]{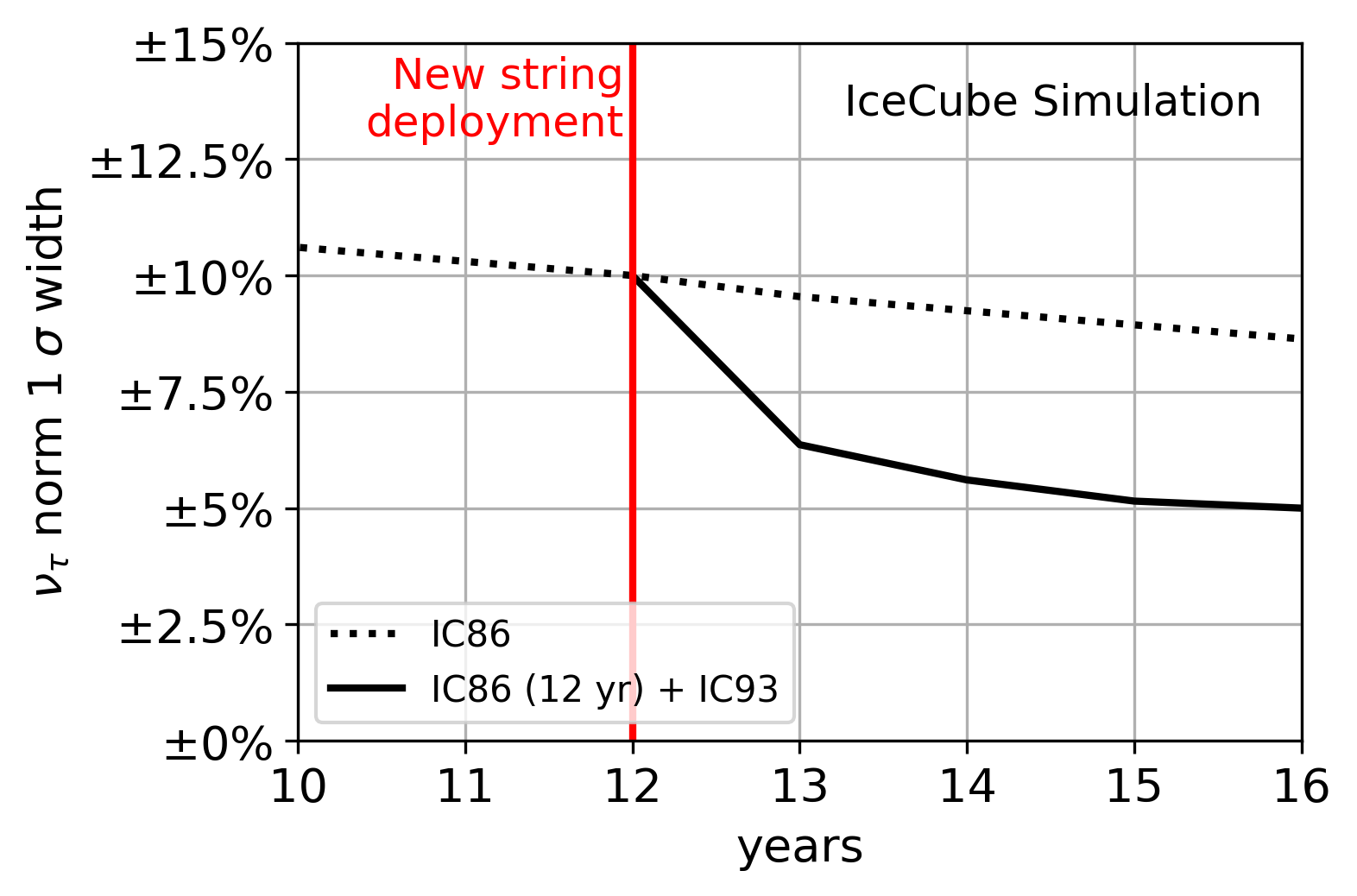}
         \caption{Width of the $1\sigma$ CL as a function of livetime}
         \label{fig:x}
     \end{subfigure}
        \caption{Sensitivity to the norm of the $\nu_\tau$ unitarity breaking parameter for the scenario with DeepCore only (dashed line) and with the Upgrade included (solid lines).}
        \label{fig:nutau_live}
\end{figure}

This analysis shows how well we can constrain the unitarity of the PMNS matrix in the tau sector by scaling the amount of $\nu_\tau$ appearance. More information about how IceCube measures $\nu_\tau$ appearance can be found in \cite{IceCube:2019nutauprd}.
Figure~\ref{fig:nutau} compares the sensitivity to the $\nu_{\tau}$ normalization with and without IceCube Upgrade. With the 3 years of data including Upgrade strings, the uncertainty can be almost reduced by a factor of two. To illustrate the evolution of this sensitivity, Fig.~\ref{fig:nutau_live} shows the $1\sigma$ uncertainty on the $\nu_\tau$ normalization as a function of the detector livetime. The new instrumentation will significantly improve IceCube's ability to constrain this parameter.

\subsection{Neutrino Mass Ordering}

\begin{figure}[h]
     \centering
     \begin{subfigure}[b]{0.4\textwidth}
         \centering
         \includegraphics[width=1.\textwidth]{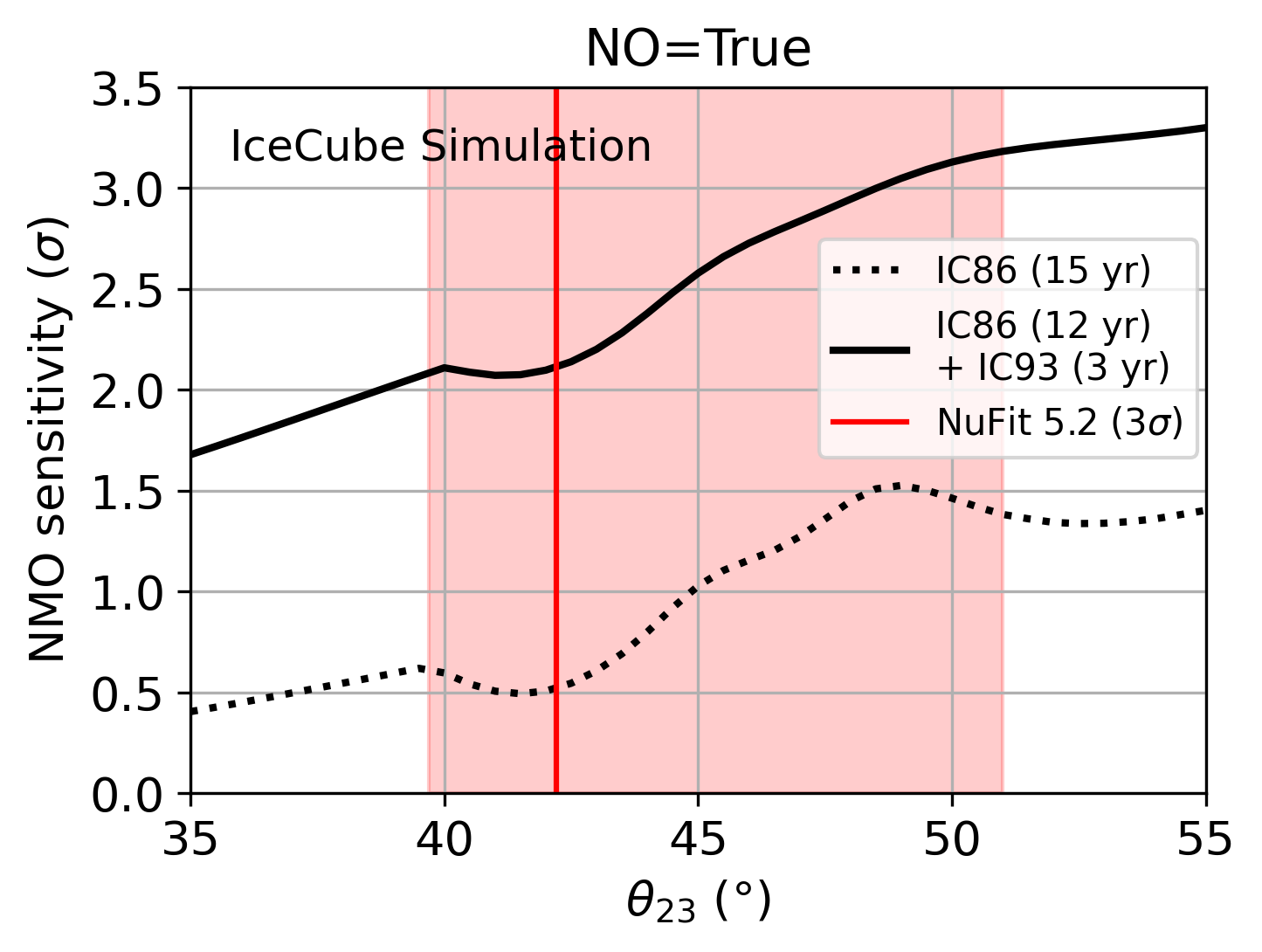}
         \label{fig:NMO_t23_NO}
     \end{subfigure}
     \hspace{0.03\textwidth}
     \begin{subfigure}[b]{0.4\textwidth}
         \centering
         \includegraphics[width=1.\textwidth]{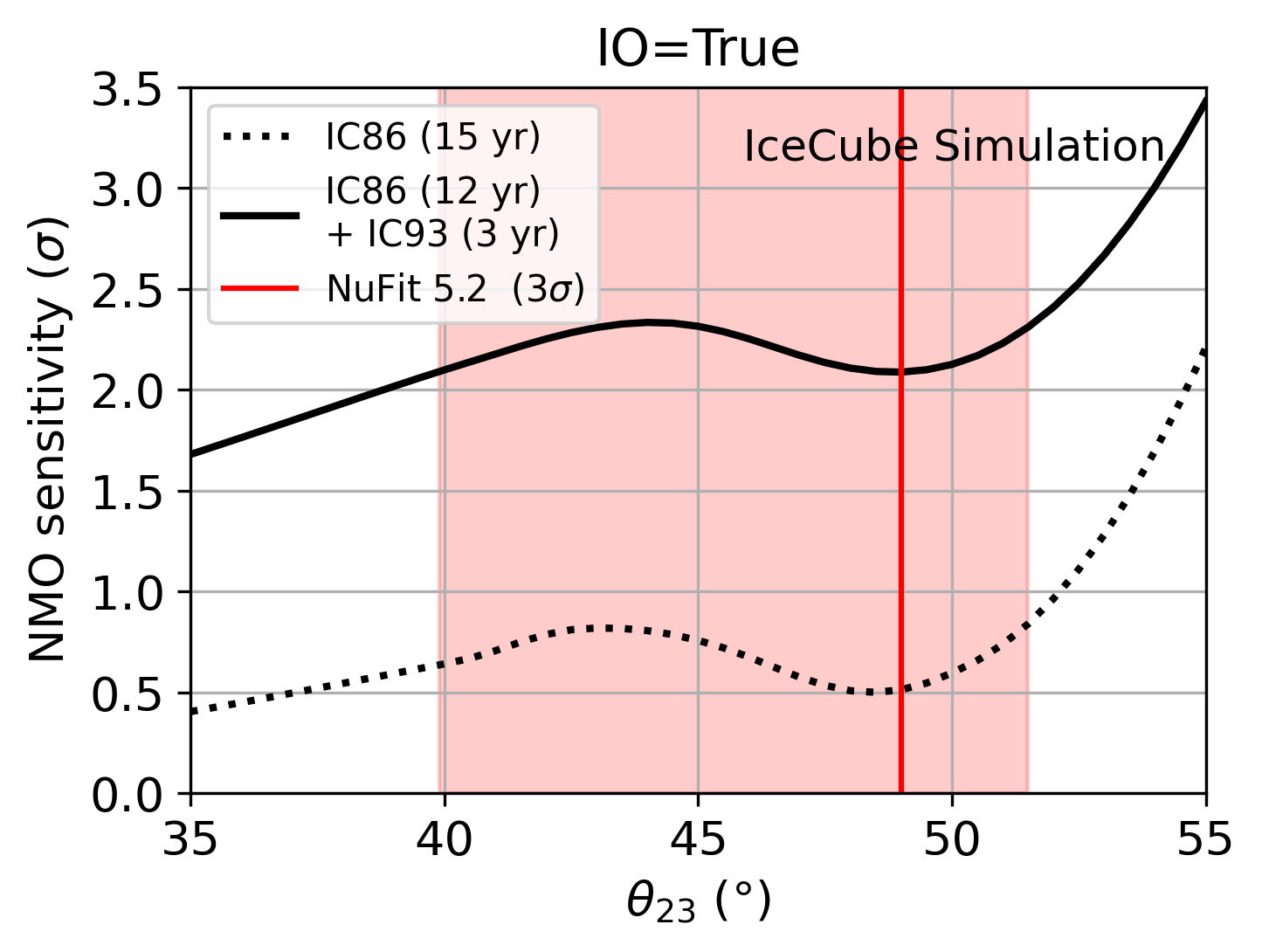}
         \label{fig:NMO_t23_IO}
     \end{subfigure}
     \vspace{-1.5em}
        \caption{Sensitivity to the NMO as a function of $\theta_{23}$ for the scenario with DeepCore only (dashed line) and with the Upgrade included (solid lines). The left panel assumes a true NO, while the right panel assumes a true IO. The red line and band indicates the current NuFit best-fit value and $\pm 3\sigma$ uncertainty on $\theta_{23}$.}
        \label{fig:NMO_t23}
\end{figure}

\begin{figure}[H]
     \centering
     \begin{subfigure}[b]{0.4\textwidth}
         \centering
         \includegraphics[width=1.\textwidth]{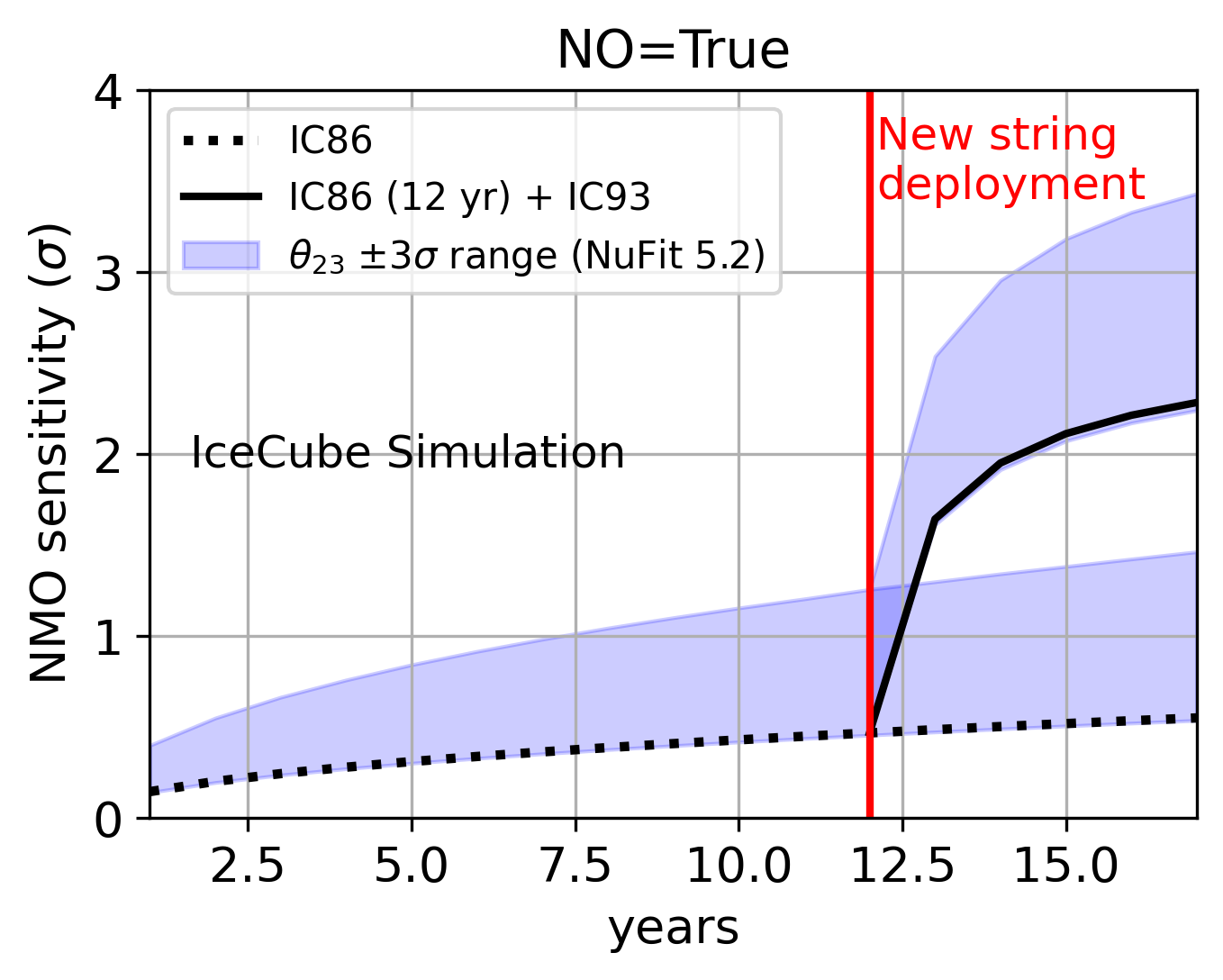}
         \label{fig:NMO_live_NO}
     \end{subfigure}
     \hspace{0.03\textwidth}
     \begin{subfigure}[b]{0.4\textwidth}
         \centering
         \includegraphics[width=1.\textwidth]{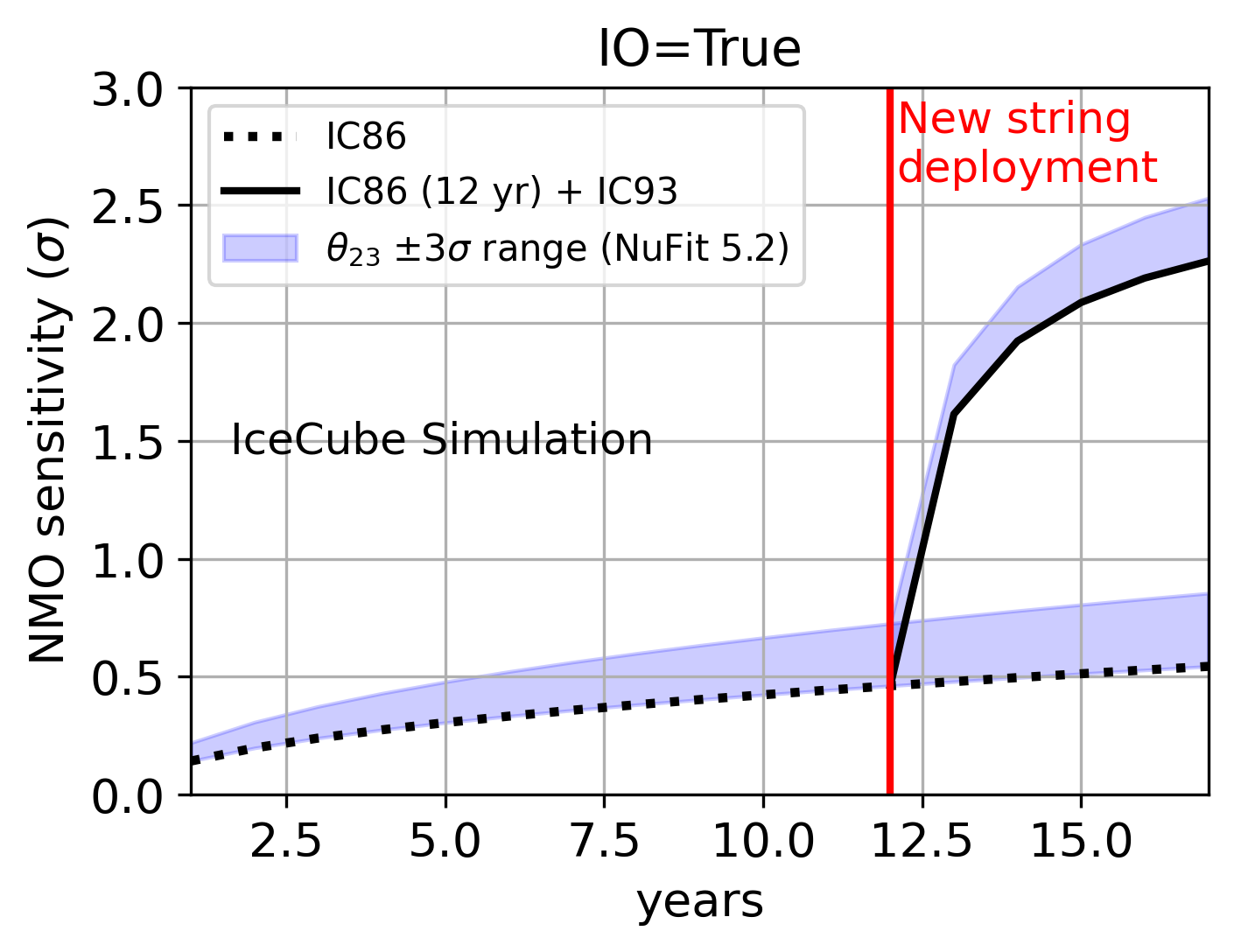}
         \label{fig:NMO_live_IO}
    \end{subfigure}   
    \vspace{-1.5em}
        \caption{Sensitivity to the NMO as a function of livetime for the scenario with DeepCore only (dashed line) and with the Upgrade included (solid lines). The left panel assumes a true NO, while the right panel assumes a true IO. While the black lines assume the current best-fit value of $\theta_{23}$ by NuFit, the blue bands show the range of possible sensitivities within the $\pm 3\sigma$ uncertainty on $\theta_{23}$.}
        \label{fig:NMO_live}
\end{figure}

Finally, the sensitivity to the neutrino mass ordering is shown, where median sensitivity is defined according to \cite{Blennow:2013oma}. More information about how IceCube determines the NMO can be found in \cite{IceCube:2019dyb}. Note that while the systematic parameters used for the NMO sensitives were updated with respect to the previous two sections, however the other analysis choices are not fully optimized in the sensitivities presented below.
IceCube's sensitivity to the NMO strongly depends on the true value of $\theta_{23}$; Fig.~\ref{fig:NMO_t23} shows the sensitivity as a function of $\theta_{23}$ for both possible orderings separately. The sensitivity is significantly improved by adding the new strings.
Figure~\ref{fig:NMO_live} shows the evolution of the NMO sensitivity over time. With the additional strings, IceCube will reach more than $2\sigma$ within a few years for any allowed value of $\theta_{23}$ and either true ordering. Assuming a true normal ordering and preferential value of $\theta_{23}$, more than $3\sigma$ sensitivities are expected within 5 years.

\section{Conclusions}
In this work we presented the first end to end, full Monte Carlo based analysis for IceCube Upgrade oscillation physics, including new, realistic methods such as the GNN event cleaning and reconstruction, as well as a new event selection pipeline.
Strong improvements over the current IceCube DeepCore detector are expected. For a livetime of 3 years of the Upgrade detector, we expect to collect 315 thousand neutrinos with the selection presented here. This combined with the 277 thousand neutrinos expected from the 12 year data taken by IceCube DeepCore up until deployment, uncertainties are reduced at the order of 20-30\% for the atmospheric neutrino oscillation parameters, around 40\% for the tau normalization and a factor of almost $4\times$ boost in NMO sensitivity is seen.
The sensitivities presented are expected to improve in the future when leveraging the additional calibration devices that will be deployed with the Upgrade, and further optimizing the triggers, event selection \& processing, and analysis choices.

\bibliographystyle{ICRC}
\bibliography{references}

%

\clearpage

\section*{Full Author List: IceCube Collaboration}

\scriptsize
\noindent
R. Abbasi$^{17}$,
M. Ackermann$^{63}$,
J. Adams$^{18}$,
S. K. Agarwalla$^{40,\: 64}$,
J. A. Aguilar$^{12}$,
M. Ahlers$^{22}$,
J.M. Alameddine$^{23}$,
N. M. Amin$^{44}$,
K. Andeen$^{42}$,
G. Anton$^{26}$
% ...
\\
\\
% ...
$^{12}$ Universit{\'e} Libre de Bruxelles, Science Faculty CP230, B-1050 Brussels, Belgium \\
% ...
$^{17}$ Department of Physics, Loyola University Chicago, Chicago, IL 60660, USA \\
$^{18}$ Dept. of Physics and Astronomy, University of Canterbury, Private Bag 4800, Christchurch, New Zealand \\
% ...
$^{22}$ Niels Bohr Institute, University of Copenhagen, DK-2100 Copenhagen, Denmark \\
$^{23}$ Dept. of Physics, TU Dortmund University, D-44221 Dortmund, Germany \\
% ...
$^{26}$ Erlangen Centre for Astroparticle Physics, Friedrich-Alexander-Universit{\"a}t Erlangen-N{\"u}rnberg, D-91058 Erlangen, Germany \\
% ...
$^{40}$ Dept. of Physics and Wisconsin IceCube Particle Astrophysics Center, University of Wisconsin{\textendash}Madison, Madison, WI 53706, USA \\
% ...
$^{42}$ Department of Physics, Marquette University, Milwaukee, WI, 53201, USA \\
% ...
$^{44}$ Bartol Research Institute and Dept. of Physics and Astronomy, University of Delaware, Newark, DE 19716, USA \\
% ...
$^{63}$ Deutsches Elektronen-Synchrotron DESY, Platanenallee 6, 15738 Zeuthen, Germany  \\
$^{64}$ Institute of Physics, Sachivalaya Marg, Sainik School Post, Bhubaneswar 751005, India \\
% ...

\subsection*{Acknowledgements}

\noindent
The authors gratefully acknowledge the support from the following agencies and institutions:
USA {\textendash} U.S. National Science Foundation-Office of Polar Programs,
U.S. National Science Foundation-Physics Division,
U.S. National Science Foundation-EPSCoR,
Wisconsin Alumni Research Foundation,
Center for High Throughput Computing (CHTC) at the University of Wisconsin{\textendash}Madison,
Open Science Grid (OSG),
Advanced Cyberinfrastructure Coordination Ecosystem: Services {\&} Support (ACCESS),
Frontera computing project at the Texas Advanced Computing Center,
U.S. Department of Energy-National Energy Research Scientific Computing Center,
Particle astrophysics research computing center at the University of Maryland,
Institute for Cyber-Enabled Research at Michigan State University,
and Astroparticle physics computational facility at Marquette University;
Belgium {\textendash} Funds for Scientific Research (FRS-FNRS and FWO),
FWO Odysseus and Big Science programmes,
and Belgian Federal Science Policy Office (Belspo);
Germany {\textendash} Bundesministerium f{\"u}r Bildung und Forschung (BMBF),
Deutsche Forschungsgemeinschaft (DFG),
Helmholtz Alliance for Astroparticle Physics (HAP),
Initiative and Networking Fund of the Helmholtz Association,
Deutsches Elektronen Synchrotron (DESY),
and High Performance Computing cluster of the RWTH Aachen;
Sweden {\textendash} Swedish Research Council,
Swedish Polar Research Secretariat,
Swedish National Infrastructure for Computing (SNIC),
and Knut and Alice Wallenberg Foundation;
European Union {\textendash} EGI Advanced Computing for research;
Australia {\textendash} Australian Research Council;
Canada {\textendash} Natural Sciences and Engineering Research Council of Canada,
Calcul Qu{\'e}bec, Compute Ontario, Canada Foundation for Innovation, WestGrid, and Compute Canada;
Denmark {\textendash} Villum Fonden, Carlsberg Foundation, and European Commission;
New Zealand {\textendash} Marsden Fund;
Japan {\textendash} Japan Society for Promotion of Science (JSPS)
and Institute for Global Prominent Research (IGPR) of Chiba University;
Korea {\textendash} National Research Foundation of Korea (NRF);
Switzerland {\textendash} Swiss National Science Foundation (SNSF);
United Kingdom {\textendash} Department of Physics, University of Oxford.

\end{document}